\documentclass[aps,prb,twocolumn,showpacs,floatfix,amsfonts,amssymb]{revtex4}

\usepackage{graphicx,graphics,color,epsfig}
\usepackage{bm}
\usepackage{amsmath}
\usepackage{amssymb}
\usepackage{amsfonts}

\newcommand{\beqa}{\begin{eqnarray}}
\newcommand{\eeqa}{\end{eqnarray}}

\begin{document}

\title{Cracking the Supersolid}

\author{Philip Phillips}
\affiliation{Department of Physics,
University of Illinois
1110 W. Green Street, Urbana, IL 61801, U.S.A.}
\author{Alexander V. Balatsky}
\affiliation{Theoretical Division and Center for Integrated Nanotechnology, Los
Alamos National Laboratory, Los Alamos, NM 87245, USA}



\maketitle


We routinely teach students in a first course in physics that the
rotational motion of a rigid body is strongly determined by its
moment of inertia.  Such an exercise would be futile in accounting for
the rotational properties of a supersolid\cite{Leggett70} in which some
 of the atoms remain still while the rest
rotate with the container. A supersolid is one of the truly
enigmatic quantum states of matter, whereby the very same atoms
exhibit simultaneously crystalline order and a superfluid stiffness.
Indeed, it is the richness of quantum mechanics that permits this
seemingly paradoxical behaviour. A series of torsional oscillator
experiments on solid He$^4$ report that anywhere from .14 to
20$\%$\cite{Chan04a,Chan04b,Chan:05,Chan06,kondo,Reppy07} of the
atoms remain still while the rest rotate with the container. The
current debate regarding the observation of a missing moment of
inertia (MMI) in solid He$^4$ centers on whether superflow and hence
supersolidity is the root cause or whether some other perhaps
non-equilibrium phase, such as a glass (which may or may not support
a superfluid stiffness), might be operative.

Of course, non-controversial instances of a missing moment of inertia or non-classical rotational inertia exist.  Consider a rotating container filled with liquid He$^4$.  At sufficiently low temperature, the angular momentum of the liquid He$^4$ disappears provided the container rotates sufficiently slowly so that vortices are not excited.  This effect, first observed by Hess and Fairbank\cite{hf} is the analogue of the Meissner effect in a superconductor and represents the benchmark test of equilibrium superfluidity. MMI in a superfluid requires that both the single-atom wavefunctions extend over the entire sample and some fraction of atoms Bose condense.  For a perfect crystalline solid, that is, one in which the number of atoms equals the number of lattice sites, neither of these conditions can be satisfied.

To rectify crystallinity with a MMI, Andreev and Lifshitz\cite{AL} and
others\cite{Leggett70,Chester67} focused on the quantum mechanical motion associated with vacancy or interstitial defects.  Such defects occur, in principle, in any solid and give rise to incommensuration. Since they are bosons, vacancies or interstitials can Bose condense at sufficiently low temperature. In such a scenario, the defects are the superfluid while the He$^4$ atoms maintain long-range crystal order.

Although the vacancy scenario has been adopted\cite{ABH} to explain
the MMI in the 2004 torsional oscillator experiments of Kim and
Chan\cite{Chan04a} (KC) on solid He$^4$, such explanations
ultimately leave a residue.  First, accurate Monte Carlo
calculations indicate that vacancy-type defects phase separate in
pure solid He$^4$ rather than form a supersolid\cite{vac,ceperley}.
Second, the experimental bounds on the number of
vacancy/interstitial defects determined by Simmons, et
al.\cite{simmons} is far lower ($0.3\%$) than the modest superfluid
fraction of 2\% seen in the early experiments of Kim and Chan. In
fact, based on the expression for the Bose condensation temperature
in the dilute regime, it is straightforward to show that vacancy
defects with a density of $.3\%$ already condense at 200mK, the
onset temperature for the MMI in the KC experiments, a further
indication that the defect scenario cannot quantitatively explain
the data. Third, Blackburn, et al.\cite{blackburn} have observed no
low-temperature anomaly in the Debye-Waller factor below the onset
temperature of the MMI in the KC experiment. Absence of such an
anomaly casts doubt on the MMI of KC being associated with a true
phase transition driven by defects.  Fourth, Rittner and
Reppy\cite{Reppy07,Sophie06} have demonstrated that the MMI signal
is acutely senstive to the quench time for solidifying the liquid.
In one extreme, they found the MMI to be absent in fully annealed
samples. This experiment is still in dispute as not all
groups\cite{null1,null2,null3} have been able to eliminate the MMI
by sample annealing. In the other, the MMI increased to an
astounding 20\% (see Fig. 1) in samples in which the solidification from the
liquid occured in less than 2 minutes\cite{Reppy07}.  A narrow
annular region which maximized the surface to volume ratio enabled
such rapid cooling of the sample and ensured a high degree of
frozen-in disorder.  This striking feature suggests that
supersolidity is not intrinsic to pure solid He$^4$. Rather, some kind
of disorder, dislocation-induced plasticity, or glassy
ordering\cite{BPG,wp06,AVB07} is the efficient cause.

Ultimately, the sharp test of supersolidity is persistent mass flow, much the way a persistent current obtains in a superconductor. The key success thus far in this regard is that of Sasaki et al.\cite{sasaki} who observed
       mass flow only in samples containing grain boundaries. However, the precise
      relationship between this experiment and
      the torsional oscillator measurements is unclear because
       mass flow was observed at
      temperatures (1.1K which is not far from the bulk
       superfluid transition temperature) vastly exceeding
       the onset temperature for MMI in the torsional oscillator experiments\cite{Chan06}, namely $T_c=0.2K$.  The team led by Beamish\cite{beamish} has looked specifically for pressure-induced mass flow
through two parts of the sample separated by a set of micron-size capillaries
 and have seen no tell-tale signature.  Relying on the fact that true superflow should exhibit a thermodynamic signature, in
  particular a diminished entropy at $T_c$, Todoshchenko\cite{todo} et al.
   measured the melting curve of He$^4$ between 10 and 320mK with
   an accuracy of $0.5\mu$bar.  They observed no deviation from the expected
   $T^4$ law due to phonons in ultra-pure samples with a He$^3$ concentration
   of $0.3$ppb.  Such an experiment is not sufficient to rule out superflow
   in the highly polycrystalline samples but is a clear indication that
   superflow is not intrinsic to the pure hcp He$^4$ but an extrinsic
   disorder-driven effect.

Hence, the question remains: Is the observation of MMI in the
torsional oscillator experiments now seen by numerous groups around
the world an example of true superflow? Short of a direct observation
of persistent mass flow, it is essential that thermodynamic measurements of the
kind performed by Todoshchenko\cite{todo} be carried out on
 the polycrystalline samples of Rittner and Reppy\cite{Reppy07}. In addition,
 neutron scattering and
 x-ray tomography measurements below 200mK on the polycrystalline samples could
offer unprecedented insight into the defect structure that enables
the observed MMI.

Ultimately, if disorder is key, then several questions arise. Most
notably, why is the signal in vycor so anomalously low? Vycor has a
surface to volume
 ratio that exceeds any of the samples used in Fig. 1 by four orders
 of magnitude, while its MMI is only 0.4$\%$.  Why?  Can grain
  boundaries account for a 20\% MMI? (see Fig.1).   Can a system in which 20\%
  of the atoms flow along grain boundaries be properly thought of
  as a crystal?  What is the quantitative  theory behind the He$^3$ enhancement of $T_c$?

\begin{figure}[tb]
\begin{center}
\includegraphics[width=7.5cm,angle=0]{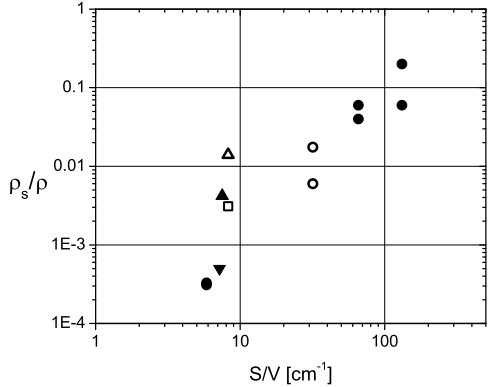}
\end{center}
\caption{ Rittner and Reppy\cite{Reppy07} noticed  that all of the
experimental values for the MMI fall onto a universal curve as a
function of the surface to volume ratio.  } \label{FigQ}
\end{figure}

In light of the disorder data, there are two classes of theoretical
proposals left standing. First, are the theories\cite{BPG,wp06} that
rely on some sort of disordered super-component present in solid
He$^4$. The essential physics of these approaches is captured by the
disordered Bose-Hubbard model \cite{MF89}.
 In fact, the only calculations\cite{wp06} that provide a quantitative
 explanation of the enhancement of $T_c$ caused by He$^3$ impurities are based on this model.  Within
  such a model, one can argue\cite{wp06} that the absence of MMI in the
  pure system arises because in the clean system, He$^4$ atoms
in an hcp lattice form a Mott insulator. That is, the atoms themselves
are the lattice sites for the Mott insulator. Disorder\cite{wp06,MF89}
   is expected to destroy the Mott state and give rise to a
   superfluid.  The precise mechanism by which this occurs, via mid-gap
    states or self-doping, is still unclear.
Second, a number of proposals have been offered for bosonic glassy
states in which MMI is obtained without ever invoking superflow
\cite{AVB07}. Torsional oscillator experiments ultimately measure
the changes in mechanical properties, oscillation period, and
damping. The connection to MMI is indirect and is not interpretation
free. One common feature of any kind of normal glassy state would be
that the mechanical response of the solid and hence the torsional
oscillator properties are changing at the onset of the glass state.
Glassy state proposals also predict a frequency-dependent decay of
the oscillation amplitude. The challenge for theory and experiment
would be to characterize bulk He$^4$ samples with enough precision
to rule in or out such non-superflow scenarios.

Fig. 1 and the dramatic enhancement He$^3$ has on $T_c$ lay plain
that the standard textbook supersolid falls short as an adequate
explanation of the experiments.  What is clear is that the true answer is hidden in the
disorder.

\end{document}